\documentclass[aps,showpacs,preprintnumbers,amsmath,amssymb,nofootinbib]{revtex4}

\usepackage{float}
\usepackage{graphics,epsfig}
\usepackage{bm}
\usepackage{slashed}
\usepackage{graphicx}
\usepackage{amsmath}
\usepackage{amsfonts}
\usepackage{amssymb}
\usepackage{color}%
\usepackage{dcolumn}
\providecommand{\U}[1]{\protect\rule{.1in}{.1in}}

\begin{document}
\baselineskip=0.6 cm \title{Thermal transport and quasi-normal modes in Gauss-Bonnet-axions theory}

\author{Xiao-Mei Kuang $^{1}$}
\email{xmeikuang@gmail.com}
\author{Jian-Pin Wu $^{2,3}$}
\email{jianpinwu@mail.bnu.edu.cn}
\affiliation{$^1$ Instituto de F\'isica, Pontificia Universidad Cat\'olica de Valpara\'iso, Casilla 4059, Valpara\'iso, Chile
\\$^2$ Institute of Gravitation and Cosmology, Department of Physics, School of Mathematics and Physics, Bohai University, Jinzhou 121013, China
\\ $^3$ Shanghai Key Laboratory of High Temperature Superconductors,
Shanghai, 200444, China}
\vspace*{0.2cm}
\begin{abstract}
\baselineskip=0.6 cm
\begin{center}
{\bf Abstract}
\end{center}
We obtain the black brane solution in arbitrary dimensional Gauss-Bonnet-axions (GBA) gravity theory.
And then the thermal conductivity of the boundary theory dual to this neutral black brane is explored.
We find that the momentum dissipation suppresses the DC thermal conductivity while it is enhanced by larger GB parameter.  The analytical and numerical results of DC thermal conductivity match very well. Also we study the effect of the momentum dissipation and the GB coupling on the AC thermal conductivity and fit the results by Drude-like behavior for low frequency. Finally, we analytical compute the quasi-normal modes (QNM) frequency of the perturbative master field in large dimensions limit. Our analytical QNM frequencies agree well with the numerical results in large enough finite dimensions.
\end{abstract}

\maketitle
\newpage
\vspace*{0.2cm}

\section{Introduction}\label{sec:Introduction}
The powerful holographic method, gauge/gravity duality\cite{Maldacena,Gubser,Witten}, provides remarkable tools to explore diversity of strongly correlated systems in condensed matter physics through studying weakly coupled bulk gravitational theory\cite{Hartnoll:2009sz,Herzog:2009xv,Iqbal:2011ae}. More recently,
to produce features of real materials, such as finite DC conductivity, we can introduce the momentum dissipation mechanism.
There are several methods involved in to achieve the goal.
One way is to introduce the so-called scalar lattice or ionic lattice,
which is implemented by periodic scalar source or chemical potential\cite{Horowitz:2012ky,Horowitz:2012gs,Donos:2012js,Ling:2013nxa}.
Also, we can implement the momentum dissipation in the holographic massive gravity framework,
which breaks the bulk diffeomorphism invariance and so that the momentum dissipates in the dual boundary field theory\cite{Vegh:2013sk,Davison:2013jba,Blake:2013owa,Blake:2013bqa,Amoretti:2014mma,Zhou:2015dha,Mozaffara:2016iwm}.
Another is holographic Q-lattice model in which the global phase of the complex scalar field breaks translational invariance\cite{Donos:2013eha,Donos:2014uba,Ling:2015exa}.
Next but not the last is the models with massless scalar fields being linear dependent on the spatial directions, which is also named as linear axions \cite{Andrade:2013gsa,Kim:2014bza,Cheng:2014tya,Ge:2014aza,Andrade:2016tbr}.

In this paper, we shall construct the black brane solution in the framework of Gauss-Bonnet-axions (GBA).
And then we numerically solve a gauge-invariant master field equation and obtain the thermal conductivity in five dimensional GBA theory.
Also, we study the quasi-normal modes (QNM) spectrum by using the large spacetime dimension (large $D$) techniques.  For a review on QNM, please refer to \cite{Konoplya:2011qq}.

The large $D$ techniques, which is proposed in Ref.\cite{Emparan:2013moa}, is
an efficient analytical tool to approximate finite $D$ results of the equations in General Relativity as a perturbative calculation in expansion $1/D$.
The main idea of the construction is that the large $D$ limit localizes the gravitational field of the black hole in a near horizon region where the gravitational potential is very steep. This well-defined near horizon region splits the QNM spectrum into two different sets, decoupling modes and non-decoupling modes. The former modes  squeezing only in the near horizon region,
are normalizable states to all orders of $1/D$  and are sensitive to different black holes and capture specific properties.
While the latter modes shared by many black holes can survive in the whole region and they are non-normalizable near horizon\cite{Emparan:2014cia,Emparan:2014aba,Emparan:2015rva}, so people are usually not interesting in the non-decoupled modes.  The large $D$ method has been applied to study the (in)stability of black holes\cite{Emparan:2014jca,Tanabe:2015isb,Chen:2016fuy}.
In holographic framework, the large $D$ method was used in \cite{Andrade:2015hpa} to pioneer to explore the analytical Drude behaviour beyond the hydrodynamic regime.
The authors focused on the normalizable decoupling modes to analytically compute the QNM of the master filed and  AC thermal conductivity which agree well with the numerical results in the large $D$ dimensional neutral black hole in Einstein-Axion theory.

The remaining of the paper is organized as follows. In section \ref{sec-background}, we solve the equations of motion derived from the GBA action and present the black brane solutions in $D=n+3 (n\geqslant 2)$ gravitational theory. Then we show the fluctuation equations of the neutral black hole  in section \ref{sec-pertur}.
In section \ref{Thermal conductivity}, we study the thermal conductivity in GBA theory.
Finally, in section \ref{Quasi-normal modes}, we apply the large $D$ method to analytically compute the QNM of the master perturbative field and compare the result with numerical one.
The last section is our conclusion and discussion.

\section{Gauss-Bonnet-axions theory}\label{sec-background}
We are interested in a specific thermal state, which
holographically dual to the Gauss-Bonnet-axions (GBA) theory
 \begin{equation}
S=\frac{1}{2\kappa^2}\int d^{n+3}x \sqrt{-g}\Big(R-2\Lambda+\frac{\alpha}{2}\mathcal{L}_{GB}-\frac{1}{2}\sum_{I=1}^{n+1}(\partial{\psi_I})^2\Big)\label{action},
\end{equation}
where $\psi_I$ are a set of axionic fields,
$2\kappa^2 = 16 \pi G_{n+3}$ is the $n+3$ dimensional gravitational coupling constant
and $\Lambda=-(n+1)(n+2)/2L^2$ is the cosmological constant.
$\alpha$ is the GB coupling constant and
\begin{equation}
\mathcal{L}_{GB}=\left(R_{\mu\nu\rho\sigma}R^{\mu\nu\rho\sigma}-4R_{\mu\nu}R^{\mu\nu}+R^2\right).
\end{equation}
In what follows, we shall set $L=1$.

With the action, the equations of motion are easily obtained as
 \begin{eqnarray}\label{eq-eom}
&&\nabla_{\mu}\nabla^{\mu}\psi_{I}=0,\nonumber\\
&&R_{\mu\nu}-\frac{1}{2}g_{\mu\nu}\Big(R+(n+1)(n+2)+\frac{\alpha}{2}(R^2-4R_{\rho\sigma}R^{\rho\sigma}+R_{\lambda\rho\sigma\tau}R^{\lambda\rho\sigma\tau})\Big)
\nonumber\\
&&+\frac{\alpha}{2}\left(2RR_{\mu\nu}-4R_{\mu\rho}R_\nu^{~\rho}-4R_{\mu\rho\nu\sigma}R^{\rho\sigma}+2R_{\mu\rho\sigma\lambda}R_\nu^{~\rho\sigma\lambda}\right)
-\sum_{I=1}^{n+1}\left(\frac{1}{2}\partial_{\mu}{\psi_I}\partial_{\nu}{\psi_I}-\frac{g_{\mu\nu}}{4}(\partial{\psi_I})^2\right)=0.
 \end{eqnarray}
We take the form of the scalar fields linearly depending on the $n+1$ spatial direction $x^a$ as \footnote{In general, the linear combination form of the scalar fields are $\psi_I=\beta_{Ia}x^a$. Then defining a constant $\beta^2\equiv \frac{1}{n+1}(\sum_{a=1}^{n+1}\sum_{I=1}^{n+1}\beta_{Ia}{\beta_{Ia}})$ with the coefficients satisfying the condition $\sum_{I=1}^{n+1}\beta_{Ia}{\beta_{Ib}}=\beta^2\delta_{ab}$, we will get the same black hole solution. Since there is rotational symmetry on the $x^a$ space, we can choose $\beta_{Ia}=\beta\delta_{Ia}$ without loss of generality.}
 \begin{equation}
 \psi_I=\beta\delta_{Ia}x^a,
 \end{equation}
which is responsible for the momentum dissipation in the dual field theory.
And then, a homogeneous and isotropic neutral black brane solution is admitted
 \begin{equation}\label{eq-metric}
ds^2=-f(r)dt^2+\frac{1}{f(r)}dr^2+\frac{r^2}{L_e^2}dx^a dx^a,
 \end{equation}
where, after defining $\hat{\alpha}=n(n-1)\alpha/2$,
 \begin{eqnarray}
 \label{eq-fr}
f(r)=\frac{r^2}{2\hat{\alpha}}\left(1-\sqrt{1-4\hat{\alpha}\left(1-\frac{r_h^{n+2}}{r^{n+2}}\right)
+\frac{2\hat{\alpha}}{r^2}\frac{L_e^2\beta^2}{n}\left(1-\frac{r_h^{n}}{r^{n}}\right)}\right).
 \end{eqnarray}
Here $r_h$ satisfying $f(r_h)=0$ is the black brane horizon. The GB coupling parameter $\hat\alpha$ is constrained by no negative energy fluxes condition and causality of  the dual CFT into the range\cite{Ge:2009eh,Camanho:2009vw,Buchel:2009sk}\footnote{It was shown in \cite{Camanho:2014apa} that the constraint in the higher derivative coupling coming from causality issues is much more severe, especially in a weakly coupled theory. And later in \cite{DAppollonio:2015fly}, the authors pointed out that the causality violations can be cured by considering the  Regge behavior. The preciser causality constraint of higher correction coupling is worthy further investigated.}
\begin{equation}\label{constraint}
-\frac{n(3n+8)}{4(n+4)^2}\leqslant \hat\alpha \leqslant \frac{n(n-1)(n^2+3n+8)}{4(n^2+n+4)^2}.
\end{equation}
Note that this result is given in the case without the axionic fields.
It may become more complicate due to the introduction of axionic fields \cite{Wang:2016vmm} and we shall address these problems in future. In this paper, we will take the constraint \eqref{constraint}.

Then, via the standard method, the Hawking temperature of the black brane is
 \begin{equation}
T=\frac{f'(r_h)}{4\pi}=\frac{1}{4\pi}\left((n+2)r_h-\frac{L_e^2\beta^2}{2r_h}\right).
 \end{equation}
And the entropy density of horizon is
 \begin{equation}
s=\frac{r_h^{n+1}}{4G_{n+3}}.
 \end{equation}
Near the UV boundary $r\rightarrow\infty$,
\begin{equation}
f(r)\sim\frac{1-\sqrt{1-4\hat{\alpha}}}{2\hat{\alpha}}r^2.
\end{equation}
So the effective  asymptotic AdS radius is
\begin{eqnarray}
L^2_{\rm e}=\frac{2\hat{\alpha}}{1-\sqrt{1-4\hat{\alpha}}}
\to  \left\{
\begin{array}{rl}
1   \ , &  \text {for} \ \hat{\alpha} \rightarrow 0 \\
\frac{1}{2}  \ , &  \text{ for} \ \hat{\alpha} \rightarrow \frac{1}{4}
\end{array}\right.
\,.
\end{eqnarray}
The Einstein limit is obtained by taking  $\hat{\alpha}\rightarrow0$, in which the
gravitational background recovers  the solution addressed in\cite{Andrade:2013gsa}.
In addition, it is worth to point out that  at  zero temperature, the near horizon geometry is $AdS_2\times \mathbb{R}^{n+1}$
with the $AdS_2$ radius $L_2=\frac{1}{n+2}$.
Note that to have a unit velocity of light, the metric component $g_{ii}$ in Eqs. (\ref{eq-metric}) and (\ref{eq-fr})
are dependent of GB coupling parameter $\hat{\alpha}$.
They are somewhat different from one presented in \cite{Cheng:2014tya} where $g_{ii}=r^2/L^2$ is independent of GB coupling $\hat{\alpha}$.
This requirement shall result in different conclusion as we see later.

\section{Linearized perturbative equations}\label{sec-pertur}
To study the heat transport, we turn on the following consistent linearized perturbation about the background \eqref{eq-metric} as
\begin{equation}\label{eq:pertur}
	\delta g_{tx} = e^{- i \omega t} r^2 h_{tx}(r) , \qquad \delta \psi_1 =  e^{- i \omega t} S(r)/\beta.
\end{equation}
And then the equations of motion can be evaluated as
\begin{eqnarray}
\label{eq-htx}
&&i\omega\left(2\hat{\alpha}-\frac{r^2}{f}\right)h_{tx}'+S'= 0, \\
\label{eq-S}
&&S''+\left(\frac{n+1}{r}+\frac{f'}{f}\right)S'+\frac{\omega^2}{f^2}S-\frac{i\omega L_e^2\beta^2}{f^2}h_{tx}= 0,
\end{eqnarray}
which govern the dynamics of the perturbations.
The prime in equations above denotes the derivative to the radius coordinate $r$.
Near the boundary $r\rightarrow\infty$, the behavior of these fields is
\begin{eqnarray}
&&h_{tx}=h_{tx}^{(0)}+\frac{h_{tx}^{(n+2)}}{r^{n+2}}+\cdot\cdot\cdot,\label{eq-BIhtx}\\
&&S=S^{(0)}+\frac{(\omega^2 S^{(0)}-i\omega L_e^2\beta^2 h_{tx}^{(0)})/2n}{r^{2}}+\cdot\cdot\cdot.\label{eq-BI-S}
\end{eqnarray}
Note that for even $n$, the above behaviors should have extra logarithms terms due to the Weyl anomaly appearing in the even boundary dimensions.
To solve the equations of motion (\ref{eq-htx}) and (\ref{eq-S}),
the purely ingoing conditions for the perturbations shall be imposed near the horizon as
\begin{eqnarray}
&&h_{tx}\sim h_1(r-r_h)^{-i\omega/4\pi T},\label{eq-Bhtx}\\
&&S\sim S_1(r-r_h)^{-i\omega/4\pi T}\label{eq-B-S}.
\end{eqnarray}

It is convenient to package the linearized equations (\ref{eq-htx}) and (\ref{eq-S}) into a gauge invariant mast field equation
\begin{equation}\label{eq-eomPhi}
\Phi''+\left(\frac{f'}{f}+\frac{n-1}{r}\right)\Phi'+
\left(\frac{\omega^2}{f^2}+\left(\frac{nf'}{r f}-\frac{2n}{r^2}\right)-\frac{L_e^2\beta^2}{f(r^2-2\hat{\alpha} f)}\right)\Phi=0\,,
\end{equation}
where
\begin{equation}\label{eq-definePhi}
\Phi(r)=\frac{r f(r)S'(r)}{i\omega}\,.
\end{equation}
Near the horizon, the master equation satisfies the ingoing conditions
\begin{equation}\label{eq-PhiH}
\Phi \sim (r-r_h)^{-i\omega/4\pi T}
\end{equation}
where we set the regular coefficient to be unit.
Near the boundary, we have
\begin{equation}
\Phi=  \left\{
\begin{array}{ll}
\Phi^{(0)}+\frac{\Phi^{(n)}}{r^n}+\cdot\cdot\cdot,   \ , &  \text {for} \ n=odd \\
\Phi^{(0)}+\frac{\Phi^{(n)}}{r^n}+\frac{L_e^{2n} \omega^n\Phi^{(0)}}{N r^{n}}\log{r}+\cdot\cdot\cdot \ , &  \text{ for} \  n=even
\end{array}\right.
\end{equation}
where the dots denote the higher order terms and $N$ depends on the dimension $n$.
Thus, $\Phi^{(0)}$ and $\Phi^{(n)}$ can be treated as the source and thermal current function of the master field $\Phi$, respectively.
And the Green function which is related with the thermal conductivity $\kappa(\omega)$ is
\begin{eqnarray}\label{eq-kappa}
\kappa(\omega)=\frac{i G(\omega)}{\omega T}=  \left\{
\begin{array}{ll}
\frac{1}{i \omega T}\frac{n \Phi^{(n)}}{L_e^{n+1} \Phi^{(0)}}   \ , &  \text {for} \ n=odd \\
\frac{1}{i\omega T}\left(\frac{n \Phi^{(n)}}{L_e^{n+1} \Phi^{(0)}}-\frac{L_e^{n-1}\omega^n}{N}\right) \ , &  \text{ for} \  n=even
\end{array}\right.
\,.
\end{eqnarray}
Compared with the asymptotic behaviors of fields $h_{tx}$ and $S$ (\eqref{eq-BIhtx} and \eqref{eq-BI-S}) and
\eqref{eq-definePhi}, we have
\begin{eqnarray}
\Phi^{(0)}=\frac{i\omega S^{(0)}+L_e^2\beta^2 h_{tx}^{(0)}}{n L_e^2},~~~~\Phi^{(n)}=\sqrt{1-4\hat{\alpha}}(n+2)h_{tx}^{(n+2)}.
\end{eqnarray}
And then the thermal conductivity can also read off from the boundary behavior of the perturbation $h_{tx}$ and $S$.

Before proceeding, to simplify our problems, we shall redefine the bulk parameters as
\begin{equation}\label{transformation}
\tilde{\alpha}=n \hat{\alpha},~~~~~\tilde{\beta}=\frac{\beta}{\sqrt{n}}.
\end{equation}
Furthermore, in the calculation of thermal conductivity,
we shall use $\hat{\beta}\equiv\tilde{\beta}/T$,
which is the only one scaling-invariant quantity of the black brane (\ref{eq-metric}) with (\ref{eq-fr}) for a given GB parameter.
Also we set $r_h=1$ in numerical calculation.

\section{Thermal conductivity}\label{Thermal conductivity}
\subsection{DC thermal conductivity}

\begin{figure}[h]
  \centering
  \includegraphics[width=0.4\textwidth]{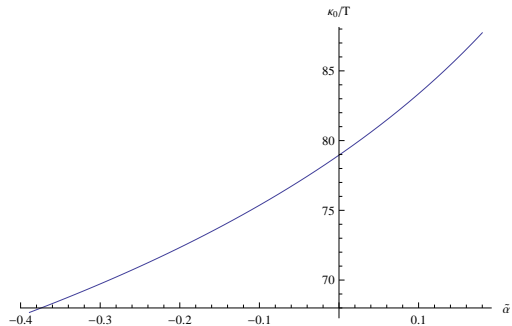}
\caption{\label{fig-alpha-kappa0} DC thermal conductivity $\kappa_0$ as a function of the GB parameter $\tilde{\alpha}$ with fixed $n=2$ and $\tilde{\beta}=1$.}
\end{figure}

In this subsection, we study the DC thermal conductivity.
Following the method outlined in \cite{Donos:2014cya,Cheng:2014tya},
we analytically derive the dimensionless DC thermal conductivity as
\begin{eqnarray}
\label{kappa0}
\kappa_0/T=\frac{(4\pi)^2 r_h^{n+1}}{n L_e^2{\tilde{\beta}}^2}.
\end{eqnarray}
Now we set $r_h=1$ and summarize its features as follows.
The mathematical details can be found in Appendix \ref{appendix}.
\begin{itemize}
 \item Form Eq. (\ref{kappa0}), it is easily observed that when $\tilde{\beta}$ goes to zero, $\kappa_0$ is divergent,
 which originates from the translational invariance and the momentum is conserved.
 When the momentum dissipates, i.e., $\tilde{\beta}$ is finite, $\kappa_0/T$ becomes finite.
 In particular, for fixed geometry parameters $n$ and $\tilde\alpha$, $\kappa_0/T$ decreases as $\tilde{\beta}$ increases.

  \item  FIG. \ref{fig-alpha-kappa0} exhibits $\kappa_0/T$ as a function of the GB parameter $\tilde{\alpha}$ for
  fixed $\tilde\beta$ and $n$. It indicates that the DC thermal conductivity is enhanced by positive GB gravity while is suppressed by negative GB coupling.
  It is worthwhile to note that, our observation is different from that shown in\cite{Cheng:2014tya} where the  $\kappa_0$ is independent of the GB coupling.
  The reason is the same as we have emphasized that to have unite speed of light, our metric component should be $g_{ii}=r^2/L_e^2$,  which depend on $\tilde{\alpha}$.

  \item  $\kappa_0/T$ decreases with the increase of the spacetime dimensions.
\end{itemize}
Next we turn to the study of AC thermal conductivity.

\subsection{AC thermal conductivity}

\begin{figure}[h]
\center{
\includegraphics[scale=0.6]{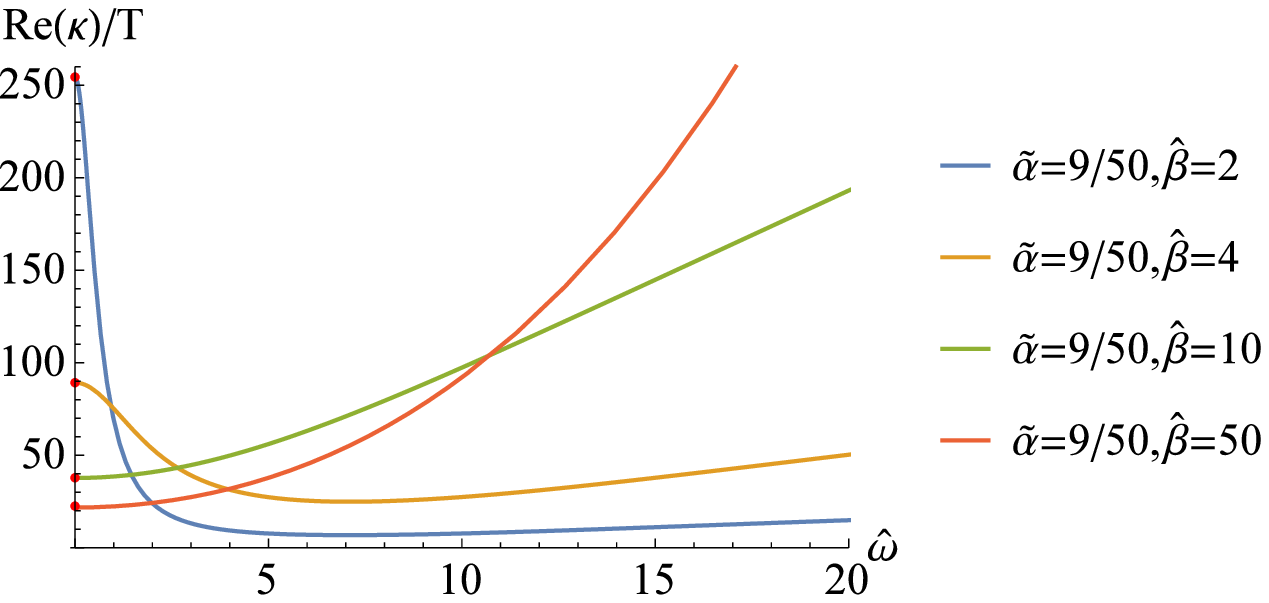}\hspace{0.5cm}
\includegraphics[scale=0.6]{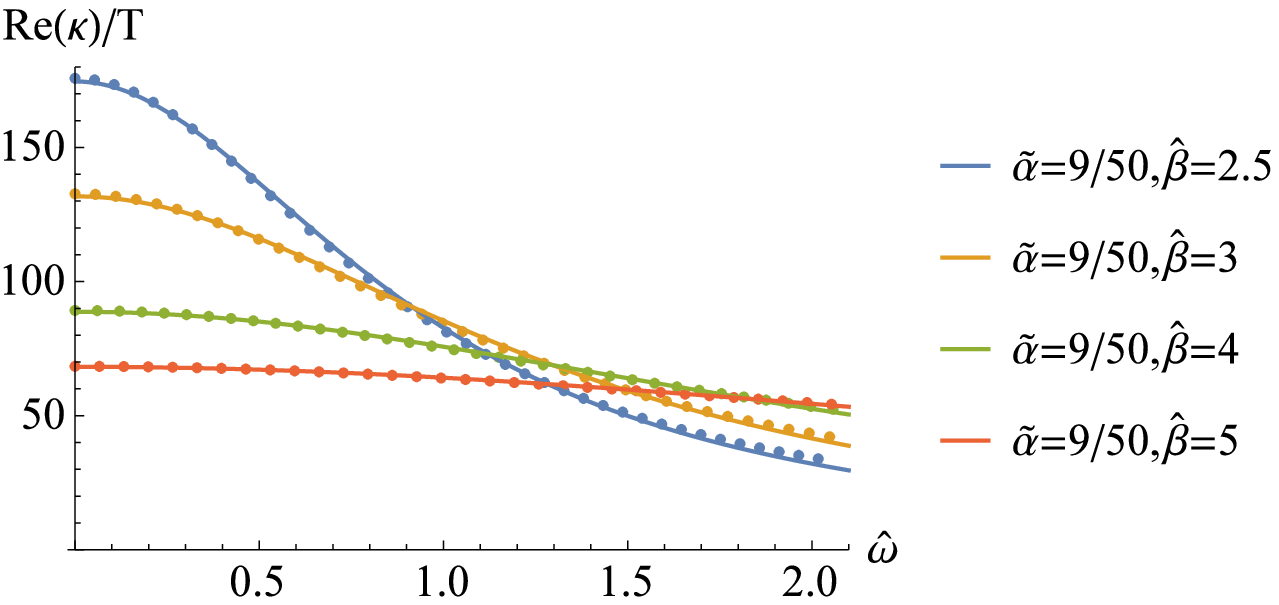}\\
\caption{\label{fig_alpha_9o50} AC thermal conductivity $\kappa(\omega)$ with fixed $\tilde{\alpha}=9/50$ for different $\hat{\beta}$.
The red dots at zero frequency in left plot are the analytic DC values calculated by (\ref{kappa0}).
Right plot: The dots are the numerical results while the solid lines are fitted by (\ref{drudeKappa}).}}
\end{figure}

\begin{figure}[h]
\center{
\includegraphics[scale=0.6]{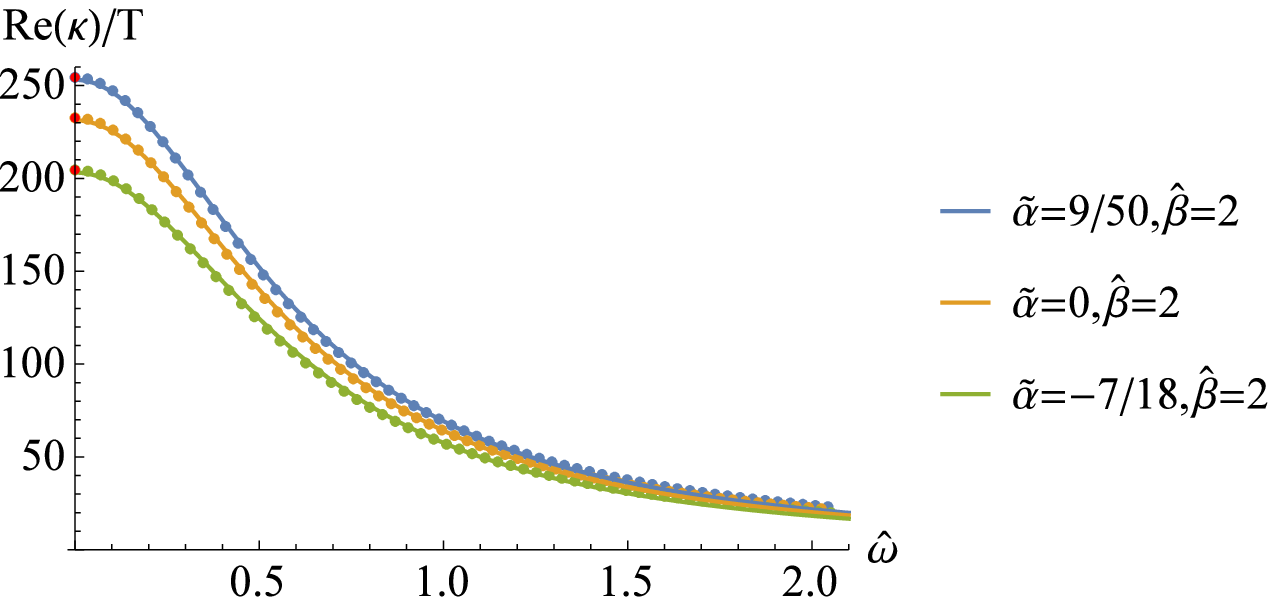}\hspace{0.5cm}
\includegraphics[scale=0.6]{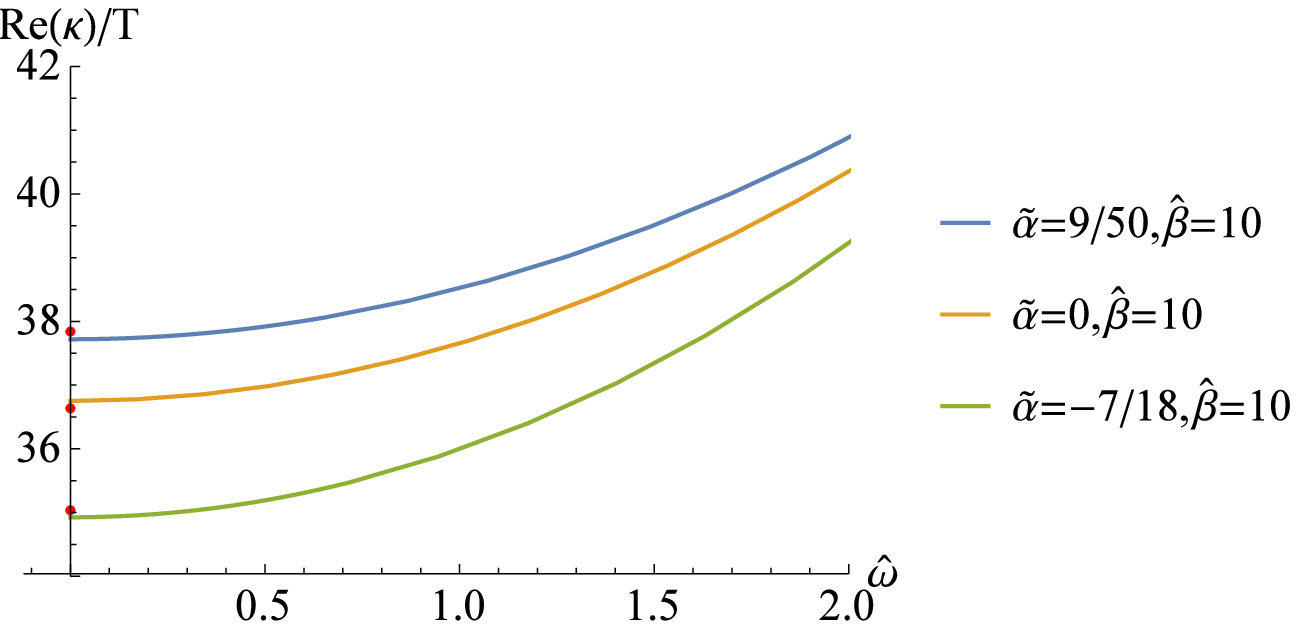}\\
\caption{\label{fig_beta_2_10} AC thermal conductivity $\kappa(\omega)$ with different $\tilde{\alpha}$ for fixed $\hat{\beta}$.
Left plot is for $\hat{\beta}=2$, in which the dots are the numerical results while the solid lines are fitted by (\ref{drudeKappa}).
Right plot is for $\hat{\beta}=10$.}}
\end{figure}

In this subsection, we study the AC thermal conductivity in five dimensional GBA gravity theory,
i.e., $n=2$, in which $\tilde\alpha$ is $-7/18\leqslant \tilde\alpha \leqslant 9/50$.
FIG.\ref{fig_alpha_9o50} shows the dissipation effect ($\hat{\beta}$) on the thermal conductivity in GBA gravity theory,
while FIG.\ref{fig_beta_2_10} exhibits the GB coupling effect on the thermal conductivity for small momentum dissipation (left plot in FIG.\ref{fig_beta_2_10})
and large momentum dissipation (right plot in FIG.\ref{fig_beta_2_10}) respectively.

Firstly, as a quick check on the consistency of our numerics,
we denotes the DC thermal conductivity analytically calculated by Eq. \eqref{kappa0} (red dots)
in FIG.\ref{fig_alpha_9o50} and FIG.\ref{fig_beta_2_10}.
They match very well with the numerical results.

And then, we focus on the momentum dissipation effect.
Since the momentum dissipates, we have finite DC thermal conductivity in GBA gravity theory
as one in Schwarzschild-axions (SA) theory \cite{Davison:2014lua,Andrade:2015hpa}.
For small momentum dissipation,
the AC thermal conductivity exhibits a Drude-like peak at low frequency.
With the increase of $\hat{\beta}$, the peak gradually becomes a valley,
which indicates a crossover from coherent to incoherent phase.
Quantitatively, for small $\hat{\beta}$, we can fit the low frequency AC thermal conductivity
in terms of the Drude-like formula (right plot in FIG.\ref{fig_alpha_9o50}),
\begin{equation}\label{drudeKappa}
\kappa(\omega)=\frac{K\tau}{1-i\omega\tau}
\end{equation}
where $K$ is a constant and $\tau$ is the relaxation time.
The corresponding fitting results of $K$ and $\tau$ with $n=2$ and $\tilde\alpha=9/50$ are listed in TABLE \ref{table-fit-beta}.

\begin{table}[h]
\center{
\begin{tabular}{|c|c|c|c|c|c|}\hline
$\hat{\beta}$&$2$&$2.5$&$3$&$4$&$5$\\\hline
$K$&$155.296$&$165.737$&$178.74$&$213.90$&$271.15$\\\hline
$\tau$&$1.630$&$1.053$&$0.737$&$0.414$&$0.251$\\\hline
 \end{tabular}
\caption{\label{table-fit-beta} Fitting parameters for different $\hat{\beta}$ with fixed $n=2$
and $\tilde{\alpha}=9/50$.}}
\end{table}
\begin{table}[h]
\center{
\begin{tabular}{|c|c|c|c|}\hline
$\tilde{\alpha}$&$9/50$&$0$&$-7/18$\\\hline
$K$&$155.28$&$143.28$&$127.76$\\\hline
$\tau$&$1.629$&$1.614$&$1.589$\\\hline
 \end{tabular}
\caption{\label{table-fit-alpha} Fitting parameters for different $\tilde{\alpha}$ with fixed $n=2$
and $\hat{\beta}=2$.}}
\end{table}

Moreover, we explore the effect of GB parameter on the AC thermal conductivity (FIG.\ref{fig_beta_2_10}).
For small $\hat{\beta}$, the peak at low frequency conductivity decreases with the decrease of GB coupling parameter $\tilde\alpha$
and the relaxation time $\tau$ decrease (TABLE \ref{table-fit-alpha}).
While for large $\hat{\beta}$, with the decrease of the GB coupling,
the valley in low frequency conductivity becomes deeper.
Nonetheless, for fixed $\hat{\beta}$ we cannot have a crossover from coherent to incoherent phase (or vice versa)
by only changing $\tilde{\alpha}$
in the allowed region of $\tilde{\alpha}$.

Note that in \cite{Andrade:2015hpa}, by matching the solutions near and far zone of bulk geometry, the authors approximately calculated the AC  thermal conductivity in large dimension limit. The Large $n$ technics, which is proposed and extensively studied in \cite{Emparan:2014cia,Emparan:2014aba,Emparan:2015rva}, provides an analytical method to perturbatively solve the above master equation by treating $1/n$ as a small parameter.
Their analytical results of AC thermal conductivity agree well with the numerical results in large dimension geometry. Here since the GB coupling complex our background solution, it is difficult to apply their skills to analytically solve our master equation in any dimension
even in the lowest order.
New skills and technics are called for and we shall explore them in future.
However,  we find that the quasi-normal modes of the master field in GBA theory can be analytically studied by the large dimension method.
And so in next section we shall work out the quasi-normal modes by large dimension method and compare it with the numerical one.

\section{Quasi-normal modes}\label{Quasi-normal modes}

We turn to study the QNM spectrum of the master field of the perturbation. Testing the perturbation by QNM to study the (in)stability of various black holes in Gauss-Bonnet theory has been addressed in \cite{Konoplya:2008ix,Cuyubamba:2016cug,Konoplya:2017ymp} and they were studied via numerical methods. Here we will investigate the QNM spectrum by the Large $n$ method \cite{Emparan:2014cia,Emparan:2014aba,Emparan:2015rva}.  Via the method, the spectrum of QNM in the large $n$ limit can be splite into decoupled modes and non-decoupled modes, of which the former are always normalisable in the near horizon geometry and unique for different black hole, while the latter are not normalisable and  have common features for many black holes.
In \cite{Andrade:2015hpa}, the authors claimed that the decoupled modes indeed controlled the  `Drude poles' of the conductivity in the boundary theory dual to the Einstein-Maxwell-axion gravity. In this subsection, we will use the Large $n$ method to analytically calculate the decoupled QNM for the master field $\Phi$ and then compare them with the numerical results.

\begin{figure}[h]
  \centering
  \includegraphics[width=0.4\textwidth]{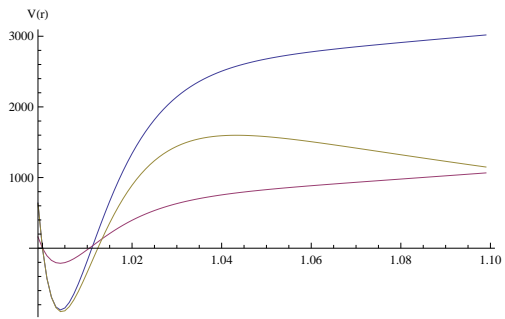}
\caption{\label{fig-r-V} The effective potential for different samples of bulk parameters.}
\end{figure}

To this end, we rewrite the master equation \eqref{eq-eomPhi} with the tortoise coordinate $dr_{*}=\frac{dr}{f(r)}$
\begin{equation}
\left(\frac{d^2}{dr_{*}^2}+\omega^2-V\right)\Psi=0\,,
\end{equation}
where we rescale $\Phi=r^{\frac{1-n}{2}}\Psi$ and the potential is
\begin{equation}
V=\frac{(n+1)f(2r f'-(n+3)f)(r^2-2\hat{\alpha} f)+4 L_e^2\beta^2r^2f}{4 r^2(r^2-2\hat{\alpha} f)}.
\end{equation}
To guarantee the existence of the decoupled QNM, the effective potential should have a negative minima \cite{Emparan:2014cia,Emparan:2014aba,Emparan:2015rva}.
FIG. \ref{fig-r-V} shows the profile of the potential for some parameters.
We see it presents a negative minima, which means that there exists the decoupled QNM of the master field.

To calculate the QNM of the decoupled mode in large $n$ limit, we introduce
\begin{equation}\label{eq-rho}
\rho=(r/r_h)^n.
\end{equation}
and expand the master field and the frequency as
\begin{equation}\label{eq-expansion}
\Phi=\sum_{k\geqslant 0}\frac{\Phi_{k}}{n^k},~~~\omega=\sum_{k\geqslant 0}\frac{\omega_{k}}{n^k}.
\end{equation}
Then, we put \eqref{eq-rho} and \eqref{eq-expansion} into the master equation \eqref{eq-eomPhi} and by holding $\tilde{\alpha}$ and $\tilde{\beta}$ fixed, expand the equation in the power of  the small quantity $1/n$. Subsequently, we can obtain series equations of motion for $\Phi_{k}$.

Before solving the equations for each
order $\Phi_{k}$ , we have to fix their boundary conditions.
In order to get the behavior of each order $\Phi_{k}$ near the horizon, we  insert the expansion \eqref{eq-expansion} into the condition  \eqref{eq-PhiH}, then solve behavior at each order
 \begin{eqnarray}\label{eq-bdyH-expansion}
&&\Phi_{0}(\rho\rightarrow1)\rightarrow1,\nonumber\\
&&\Phi_{1}(\rho\rightarrow1)\rightarrow-\frac{2i\log (\rho-1)\omega_{0}}{2-L_e^2\tilde{\beta}^2},\nonumber\\
&&\Phi_{2}(\rho\rightarrow1)\rightarrow \frac{2i \log(\rho-1)(4\omega_0+i \log(\rho-1)\omega_0^2-(2-L_e^2\tilde{\beta}^2)\omega_1)}{(2-L_e^2\tilde{\beta}^2)^2},\nonumber\\
&&\cdots\cdots\cdots\cdots
\end{eqnarray}

Near AdS boundary, the decoupled mode is normalisable, so that we have the behavior
\begin{equation}\label{eq-bdy-expansion}
\Phi_{k}(\rho\rightarrow\infty)\rightarrow 0.
\end{equation}
\begin{figure}[h]
  \centering
  \includegraphics[width=0.9\textwidth]{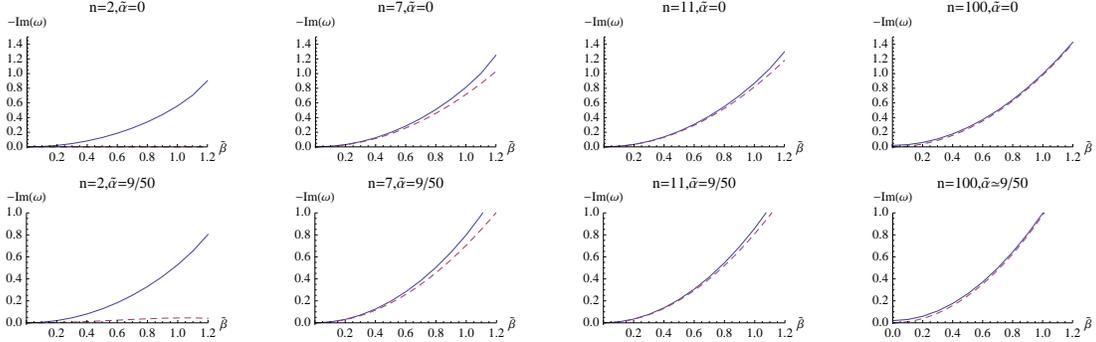}
\caption{\label{fig-w-b} Imaginary part of QNM frequency as a function of $\tilde{\beta}$ with fixed GB coupling $\tilde{\alpha}=n\hat{\alpha}$. In each plot, solid line are numerical result, while the dashed line denotes analytical result drawn via  the formular \eqref{eq-w}.}
\end{figure}
\begin{figure}[h]
  \centering
  \includegraphics[width=0.9\textwidth]{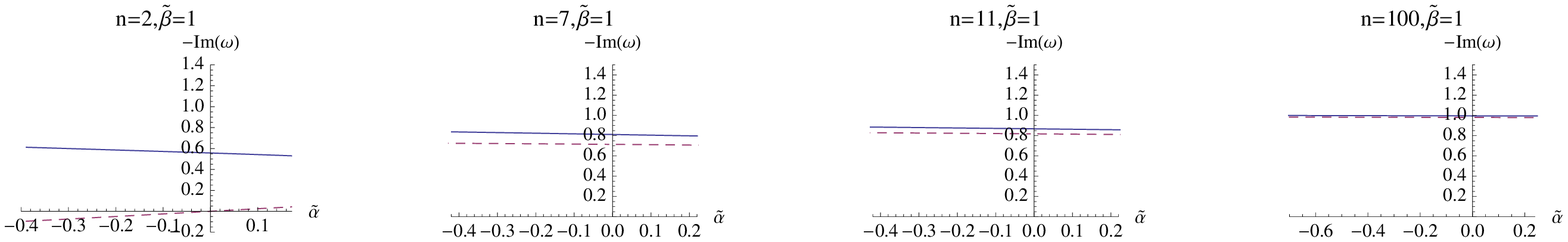}
\caption{\label{fig-w-a} Imaginary part of QNM frequency as a function of  the GB coupling $\tilde{\alpha}=n\hat{\alpha}$ with fixed axion $\tilde{\beta}$. Solid lines are numerical results, while the dashed lines are analytical results drawn via  the formular \eqref{eq-w}.}
\end{figure}
Integrating the series equation of motion order by order and considering the boundary conditions \eqref{eq-bdyH-expansion}
and \eqref{eq-bdy-expansion}, we can get the exact solution of $\Phi_0$, $\Phi_1$ and $\Phi_2$  and the frequency to the first order
 \begin{eqnarray}\label{eq-w-expansion}
\omega_0&=&-iL_e^2\tilde{\beta}^2,\nonumber\\
\omega_1&=&2iL_e^2\tilde{\beta}^2-\frac{i\tilde{\alpha}L_e^2\tilde{\beta}^2(2-L_e^2\tilde{\beta}^2)}{2}\,.
\end{eqnarray}
So that to the order $n^{-1}$, the decoupled QNM frequency for $\Phi$ is
 \begin{eqnarray}\label{eq-w}
\omega=-iL_e^2\tilde{\beta}^2\left(1+\frac{1}{n}\frac{\tilde{\alpha}(2-L_e^2\tilde{\beta}^2)-4}{2}+\textsl{O}(n^{-2})\right).
\end{eqnarray}
Note that in the Einstein limit $\tilde{\alpha}=0$, this result recovers that shown in \cite{Andrade:2015hpa}.
Due to the complex of our metric, we can only obtain the solution of $\Phi_k$ up to the second order.
Correspondingly the frequency mode we can obtained here is up to the order of $n^{-1}$.

Now we move on to compare our analytical QNM frequency \eqref{eq-w} with the numerical results to see how well the lange-n results match with the numerical results.
In FIG. \ref{fig-w-b}, we fix the scaled GB coupling $\tilde{\alpha}$ ($\tilde{\alpha}=0$ in the upper panel and $\tilde{\alpha}=9/50$ in the botton panel) and study the effect of the scaled axion momentum
$\tilde{\beta}$ on the imaginary part of QNM frequency. From the left to right plot, we increase the dimension of the background. We see that in the low dimension case, the analytical results (dashed) have a wide difference from the numerical results (solid), while as $n$ increases ($n\gtrsim 11$), the analytical and numerical results match better and better. FIG. \ref{fig-w-a} shows the imaginary part of QNM frequency
as a function of the scaled GB coupling $\tilde{\alpha}=n\hat{\alpha}$ with fixed non-vanishing axion $\tilde{\beta}=1$. It is also obvious that the
analytical and numerical results are almost consistent for large enough dimension.

\section{Conclusion and Discussion}

In this paper, we construct a new neutral black brane solution from the GBA gravity theory in any dimensional AdS spacetimes.
The thermal conductivity of the dual theory of this black brane geometry is explored.
For small momentum dissipation, the optical conductivity exhibits a Drude-like peak at low frequency region.
With the increase of the momentum dissipation,
a transition from coherent to incoherent phase happens
as has been revealed in (SA) theory \cite{Davison:2014lua,Andrade:2015hpa}.
However, we can not observe an apparent transition from coherent to incoherent phase
when we tune the GB coupling parameter only in the allowed region but fix the momentum dissipation parameter.

Also, via the large dimension technique, we analytically study the QNM of the master field which control heat transport of the dual system.
The analytical frequency of QNM agrees well with the numerical one when the dimension is large enough.
However, we would like to point out that here we only obtain the solution of mast field up to the second order as well as the frequency mode up to the order of $n^{-1}$
and we can not observe the breakdown of the perturbative expansion, which can be seen as a signature of the coherent/incoherent transition as pointed out in \cite{Andrade:2015hpa}.
It would be very interesting to improve the computation skill to solve higher order equation of the master field so that we can test the robustness observed in \cite{Andrade:2015hpa}.

In addition, we can also study the electric and heat transport by adding a gauge field term $S_M=\frac{1}{2\kappa^2}\int d^{n+3}x \sqrt{-g}\Big(-\frac{1}{4}
F_{\mu\nu}F^{\mu\nu}\Big)$ into the action \eqref{action}. An analytical charged black brane solution can be obtained as following
\begin{eqnarray}
 &&A_t(r)=\mu\left(1-\frac{r_h^n}{r^n}\right),\\
&&f(r)=\frac{r^2}{2\hat{\alpha}}\left(1-\sqrt{1-4\hat{\alpha}\left(1-\frac{r_h^{n+2}}{r^{n+2}}\right)+\frac{2\hat{\alpha}}{r^2}\left(\frac{n\mu^2r_h^n}{(n+1)r^{n}}+\frac{L_e^2\beta^2}{n}\right)\left(1-\frac{r_h^{n}}{r^{n}}\right)}\right),
 \end{eqnarray}
where $\mu$ is understood as the chemical potential of the dual field theory on the boundary.
The properties of transport of this charged black brane geometry and it related QNM shall be addressed somewhere else.

\begin{acknowledgments}
We are grateful to Kentaro Tanabe for helpful correspondence. We also thank Long Cheng, Yi Ling, Wei-Jia Li and Xiang-Rong Zheng for valuable discussions.
X. M. Kuang is supported by Chilean FONDECYT grant No.3150006.
J. P. Wu
is supported by the Natural Science Foundation of China under
Grant Nos.11305018 and 11275208, by Natural Science Foundation of Liaoning Province under
Grant Nos.201602013,
and by the grant (No.14DZ2260700) from the Opening Project of Shanghai Key Laboratory
of High Temperature Superconductors.
\end{acknowledgments}

\appendix
\section{Calculation of DC thermal conductivity in GBA theory}\label{appendix}
In order to compute the DC thermal conductivity of our background,
we closely follow the method proposed in \cite{Donos:2014cya}
and extend the result of \cite{Cheng:2014tya} to that in any dimensional background.
We consider the perturbations of the metric \eqref{eq-metric}
\begin{eqnarray}
&&g_{tx}\to t\delta h(r)+\delta g_{tx}(r),\nonumber\\
&&g_{rx}\to r^2\delta g_{rx}(r),\nonumber\\
&&\psi_1\to \psi_1+\delta \psi_1(r).
\end{eqnarray}
Then, the linearized $rx$ component of Einstein equations is
\begin{equation}\label{EErx}
\delta g_{rx}=\frac{(r^2-2\hat{\alpha}f)(r\delta h'-2\delta h)}{r^3 L_e^2\beta^2f}+\frac{\delta\psi_1'}{L_e^2\beta}
\end{equation}
and the $tx$ component is
\begin{eqnarray}\label{EEtx}
&&(r^{n-1}f-2\hat{\alpha}r^{n-3}f^2)(\delta g_{tx}+t\delta h)''+[(n-1)r^{n-2}f-2(n-3)\hat{\alpha}r^{n-4}f^2-2\hat{\alpha}r^{n-3}ff'](\delta g_{tx}+t\delta h)'\nonumber\\
&&+[2\hat\alpha r^{n-3}ff''+2(n-3)\hat{\alpha}r^{n-4}ff'-(n-1)r^{n-2}f'+2\hat{\alpha}r^{n-3}f'^2-r^{n-1}f''](\delta g_{tx}+t\delta h)=0.
\end{eqnarray}
The Einstein equation \eqref{EEtx} can be written as a total derivation, so that we can define the conserved heat current
\begin{equation}\label{heart current}
Q=(r^{n-1}-2\hat{\alpha}r^{n-3}f)[f(\delta g_{tx}+t\delta h)'-f'(\delta g_{tx}+t\delta h)].
\end{equation}
By setting $\delta h(r)=-\zeta f(r) $, we can simplify the conserved heat current as
\begin{equation}
Q=(r^{n-1}-2\hat{\alpha}r^{n-3}f)(f\delta g_{tx}'-f'\delta g_{tx}).
\end{equation}
which is time-independent.

In order to make the metric regular at the horizon, we should require the perturbation to satisfy
\begin{equation}\label{regularPertur}
\delta g_{tx}\sim r^2 f\delta g_{rx}|_{r\to r_h}-\frac{\zeta f}{4\pi T}\log(r-r_h)+\cdots
\end{equation}
and $\delta\psi_1$ to be constant at the horizon.
Since the heat current is constant in $r$ direction, we can calculate $Q$ at the horizon which gives
\begin{eqnarray}
Q=Q|_{r\to r_h}=-r^{n-1}f'\delta g_{tx}|_{r\to r_h}=\frac{\zeta (4\pi T)^2r_h^{n+1}}{L_e^2\beta^2}
\end{eqnarray}
where we have used the formulas \eqref{regularPertur} and \eqref{EErx} in the third equality. Furthermore,  the DC thermal conductivity in $x_1$ direction is
\begin{equation}\label{appenKappa0}
\kappa_0=\frac{\partial Q}{T\partial \zeta}=\frac{(4\pi)^2 T r_h^{n+1}}{L_e^2\beta^2}.
\end{equation}
And then, the dimensionless DC thermal conductivity can be written as
\begin{eqnarray}
\label{kappa0v1}
\kappa_0/T=\frac{(4\pi)^2 r_h^{n+1}}{n L_e^2{\tilde{\beta}}^2}.
\end{eqnarray}
where we have used Eq.\eqref{transformation}.

\bibliographystyle{../preprint}
\bibliography{../Bib/QuantGra.bib}

\end{document}